\documentstyle[12pt,aaspp4,psfig]{article}  
\def\be{\begin{equation}}
\def\ee{\end{equation}}

\def\lsim{\lower 2pt \hbox{$\, \buildrel {\scriptstyle <}\over
         {\scriptstyle \sim}\,$}}
\newcommand\gsim{\buildrel > \over \sim}
\begin{document}
\newcommand{\figureout}[4]{\psfig{figure=#1,width=#2,angle=#3} 
   \figcaption{#4} }

\title{Magnetar Spin-Down }

\author{Alice K. Harding, Ioannis Contopoulos and Demosthenes Kazanas}
\affil{Laboratory for High Energy Astrophysics \\
       NASA/Goddard Space Flight Center \\
       Greenbelt, MD 20771}


\begin{abstract}
We examine the effects of a relativistic wind on the spin down of
a neutron star and apply our results to the study of Soft Gamma 
Repeaters (SGRs), thought to be neutron stars with magnetic fields
$> 10^{14}$ G. We derive a spin-down formula that includes torques 
from both dipole radiation and episodic or continuous particle winds.  
We find that if SGR1806-20 puts out a continuous particle wind of 
$10^{37}\ {\rm erg}\ {\rm s}^{-1}$, then the pulsar age is consistent 
with that of the surrounding supernova remnant, 
but the derived surface dipole magnetic field is only $3 \times 10^{13}$ G, 
in the 
range of normal radio pulsars.  If instead, the particle wind flows are episodic 
with small duty cycle, then the observed period derivatives imply 
magnetar-strength fields, while still allowing characteristic ages within a 
factor of two of the estimated supernova remnant age.  Close monitoring of 
the periods 
of SGRs will be able to establish or place limits on the wind duty cycle and 
thus the magnetic field and age of the neutron star.

\end{abstract}



\section{INTRODUCTION}

There is a growing collection of pulsating high-energy sources which
occupy a unique phase-space in their combination of long ($> 5$ s), 
monotonically increasing periods and high period derivatives.  One
subgroup in this collection are the soft gamma-ray repeaters (SGRs),
transient sources that exhibit repeated bursts of soft ($\sim 30$ keV), 
short duration ($0.1$ s) $\gamma$-rays.  Bursts of average 
energy $10^{41}$ ergs repeat on irregular intervals, while giant 
bursts of energy $\sim 10^{45}$ ergs have been observed once in each
of the sources SGR 0526-66 and SGR 1900+14.  Recently, a 
period of $P = 7.47$ s has been detected from SGR 1806-20 in quiescent 
emission (Kouveliotou et al. 1998) and a period of $P = 5.16$ s 
from SGR 1900+14 in both quiescent (Kouveliotou et al. 1999) and 
giant burst emission (Hurley et al. 1999a).  Both have measured period 
derivatives around $\dot P \sim 10^{-10} \rm s\,s^{-1}$. Another group 
of sources having similar $P$ and $\dot P$ are the
anomalous X-ray pulsars (AXPs), pulsating X-ray sources with 
periods in the range $6 - 12$ s and period derivatives in the range 
$10^{-12} - 10^{-11}\rm s\,s^{-1}$ (Gotthelf \& Vasisht 1998).  These 
sources have shown only quiescent emission with no bursting behavior.  

The most plausible and commonly invoked model to explain the characteristics 
of both the SGRs and AXPs is that of a neutron star 
having a dipole magnetic field of $> 10^{14}$ G, much higher than the fields 
of ordinary, isolated pulsars and well above the quantum critical field of 
$4.4 \times 10^{13}$ G.  Such stars, known as magnetars, were 
first proposed by Duncan \& Thompson (1992), Usov (1992) and Paczynski
(1992) to account for various properties of SGRs and $\gamma$-ray 
bursts.  
All of the SGR and AXP sources are found in or near young ($\tau < 10^5$ 
yr) supernova remnants (SNR).  SGRs in addition have been associated with 
X-ray and radio plerions whose emission power far
exceeds the dipole spin-down power.  It was therefore proposed that
relativistic particle outflows from the SGR bursts (Tavani 1994, Harding 1995, 
Frail et al. 1997) or from a steady flux 
of Alfven waves (Thompson \& Duncan 1996), provide the power for  
the nebular emission.  
The existence of such a wind has been inferred indirectly by X-ray and
radio observations of the synchrotron nebula G10.0--0.3 around
SGR1806--20 (Murakami et al. 1994, Kulkarni et al. 1994)\footnote{Note that 
a revised IPN location now places the SGR source outside the core of the 
radio plerion (Hurley et al. 1999b)}. Thompson \&
Duncan~(1996) estimated that the particle luminosity from SGR1806--20
is of the order of $10^{37}\ {\rm erg}\ {\rm s}^{-1}$. 
Such energetic particle winds will also affect the spin-down  
torque of the star by distorting the dipole field structure near the 
light cylinder (Thompson \& Blaes 1998 [TB98]).  We will show, however, 
deriving a formula similar to the one given by TB98, that if relativistic wind outflow continuously dominates the spin-down of the neutron star in 
SGR 1806-20 at a level $\gsim 10^{36}\rm erg\,s^{-1}$, 
then the surface magnetic dipole field is
too low to be consistent with a magnetar model.  It is possible however that the 
wind outflows from SGR sources are episodic, lasting for a time following 
each burst that may be small compared to the time between 
bursts.  In this case, dipole radiation will dominate the spin-down
between bursts, even though wind outflow may 
dominate the average rotational energy-loss rate.  
Using a general formula for the spin-down torque that includes both dipole 
radiation and episodic particle winds, we derive the
magnetic field and characteristic age of the neutron star as a function 
of the observed $P$ and $\dot P$, the wind luminosity and wind duty cycle.
We find that the derived surface field and age can have a range of values between 
the pure dipole and pure wind cases that depend on the duty cycle of
the wind outflows, even for a constant value of the average wind 
luminosity.

\section{ENERGY LOSS FROM A WIND}

It is known that neutron star
rotation drastically distorts the magnetic field near and beyond
the light cylinder radius $R_{\rm LC}=c/\Omega$ (where $\Omega=2\pi/P$
is the neutron star angular velocity; $P$ is the period), and that
magnetic field lines crossing the light cylinder remain open all
the way to `infinity'. Field lines will also open up in the presence
of a powerful wind of particles emitted from the neutron star. Near 
the surface of the star, dipole magnetic field pressure is high enough 
to completely dominate the wind stresses. However, the magnetic 
energy density drops much faster with distance than that of a 
quasi--isotropic particle wind of kinetic luminosity $L_p$ at infinity; 
thus, ignoring a transitional region between these two regimes, beyond a 
distance of the order $r_{\rm open}\sim r_0 (B_0^2 r_0^2 c/2 L_p)^{1/4}$
magnetic field lines will be `combed out' by the wind. Here, $B_0$ is
the value of the neutron star surface dipole magnetic field and
$r_0\sim 10$~km, is the radius of the star (henceforth, we will use
small $r$ to denote spherical distances [from the center], and capital
$R$ for cylindrical distances [from the axis]). The fraction of open
field lines originates from a region of radial extent
\begin{equation}
R_{pc}\approx r_0 (r_0/r_{\rm open})^{1/2}\ll r_0
\label{polarcap}
\end{equation}
around the axis, the so-called polar cap (this estimate is obtained
for an undistorted dipole).

An aligned magnetic rotator, the  simplified geometry examined
herein, spins down (even though an aligned magnetic 
rotator in vacuum {\em does not}), because a neutron star 
magnetosphere {\em
cannot be} a true vacuum (Goldreich \& Julian~1969). 
As is discussed in Contopoulos, Kazanas \&
Fendt~(1999; hereafter CKF), a neutron star magnetosphere is
spontaneously charged in order to support the charges and
electric currents required in the realistic solution. An electric 
current also flows, in a large scale electric circuit through the 
polar cap to infinity, closing along an equatorial current sheet
discontinuity, which connects to the interface between open and closed
field lines at the light cylinder, with the circuit closing {\em across}
field lines along the polar cap. This
electric current $I_{pc}$ flows along the magnetic field lines
crossing the polar cap and generates the required
spin-down magnetic torque on the neutron star. The
neutron star spins down because the electromagnetic torque generated at
the surface of the polar cap opposes its rotation.
Inside the neutron star surface, this current
flows horizontally towards the edge of the polar cap, where it flows
out in a current sheet along the interface between open and closed
field lines. The electric current flowing through the polar cap is (to a
good approximation) distributed as $I\propto \Psi(2-\Psi/\Psi_{pc})$,
where $\Psi$ is the total amount of magnetic flux contained inside
cylindrical radius $R$, and $\Psi_{pc}$ is the total amount of
magnetic flux which opens up to infinity. This is an exact expression
for a magnetic (split) monopole, and CKF showed that it remains
approximately valid even for a dipole. Since on the neutron star surface
$\Psi\propto R^2$ when $R < R_{pc}\ll r_0$,
\begin{equation}
I(R)=I_{pc}\left(\frac{R}{R_{pc}}\right)^2
\left[2-\left(\frac{R}{R_{pc}}\right)^2\right]\ .
\label{A1}
\end{equation}
When this current flows horizontally in a layer of thickness $h(R)$ across
the polar cap, an electric current density
\begin{equation}
J(R)=\frac{I_{pc}}{2\pi R h(R)}\left(\frac{R}{R_{pc}}\right)^2
\left[2-\left(\frac{R}{R_{pc}}\right)^2\right]
\label{A2}
\end{equation}
will flow horizontally, which, combined with the axial magnetic field $B_*$
threading the polar cap, generates an azimuthal Lorenz force per unit
volume, $f(R)=\frac{1}{c}J(R)\times B_0$.
Integrating $f(R)$ over the volume of the polar cap crust where
the above electric current flows horizontally, and doubling our result
to account for the two (north/south) polar caps, we obtain the total
electromagnetic torque
\begin{equation}
T \sim \frac{2}{3c}I_{pc} B_0 R_{pc}^2.
\label{torque}
\end{equation}

We present two simple, physically equivalent, estimates of the
electric current flowing in the magnetosphere.
One is to consider particles (electrons/positrons) with
Goldreich--Julian charge densities $\rho_{GJ}\approx B_0 /2\pi R_{\rm LC}$
flowing outwards at the speed of light from the polar cap. This naive
estimate gives
\begin{equation}
I_{pc}\sim \pi R_{pc}^2 \rho_{GJ} c=
\frac{B_0 r_0 c}{2}\left(\frac{r_0}{R_{\rm LC}}\right)
\left(\frac{r_0}{r_{\rm open}}\right)
\label{current}
\end{equation}
 
Another equivalent, more physical, way to estimate the total amount of
electric current flowing in the magnetosphere through the polar cap is
to make the naive (and correct) assumption that, at the distance of
the light cylinder the two magnetic field components (toroidal and
poloidal) must be of the same order of magnitude,
$B_\phi|_{\rm LC}\sim B_p|_{\rm LC}\ .$
This is indeed true in the force--free axisymmetric magnetosphere
(without wind), since the light cylinder is the Alfv\'{e}n point (Li
\& Melrose~1994).  In general, the Alfv\'{e}n point is some short
distance inside the light cylinder. When the two field components are
scaled back to the surface of the star at the edge of the polar cap,
we obtain
\begin{equation}  \label{Bpc}
B_p|_{pc}=B_p|_{\rm LC}\left(\frac{R_{\rm LC}}{r_{\rm open}}\right)^{2}
\left(\frac{r_{\rm open}}{r_0}\right)^{3}\equiv B_0\ .
\label{fieldspc}
\end{equation}
The structure
of the field is dipole--like out to $r_{\rm open}$ and monopole--like
out to the light cylinder.  From Eqn (\ref{Bpc}) and the relation 
$B_\phi|_{\rm LC}\sim B_p|_{\rm LC}$, we have
\begin{equation}
B_\phi|_{pc}=B_0\frac{r_0^3}{R_{pc}R_{\rm LC}r_{\rm open}}
\ ,{\rm and}\ {\rm therefore}
\end{equation}
\begin{equation}
I_{pc}\sim \frac{B_\phi|_{pc} R_{pc} c}{2}=
\frac{B_0 r_0 c}{2}\left(\frac{r_0}{R_{\rm LC}}\right)
\left(\frac{r_0}{r_{\rm open}}\right)\ .
\end{equation}
This is a very simple result, and shows that the two ways of looking
at the problem are equivalent. 

Using the above relation for the polar cap current in equation~(\ref{torque})
we obtain the expression for the torque $T$ associated with the 
above spin--down arguments. The corresponding energy loss rate due to the 
above torque is therefore $\dot E = T \cdot \Omega$, given more explicitly
by
\begin{equation} \label{Edot_w}
\dot E_w = T \cdot \Omega 
=\frac{B_0^2 \, r_0^6 \, \Omega^4}{3 \, c^3} \left(\frac{R_{\rm LC}}
{r_{\rm open}} \right)^2 = \dot E_D\,\left({L_p\over \dot E_D}\right)^{1/2}
\end{equation}
where we have used $L_p/4\pi c r_{\rm open}^2 = B(r_{\rm open})^2/8\pi$ 
to obtain an expression for $\dot E_w$ in terms of the wind luminosity, 
$L_p$, and dipole energy loss, $\dot E_D$.
Note that the standard dipole spin--down formula
(modulo the different numerical factor in the denominator) is modified
by the term $(R_{\rm LC}/r_{\rm open})^2$ which incorporates the
effects of the ``loading" of the magnetosphere with the outflowing
wind.  This expression, while of similar functional form, differs 
from that of TB98 because of a normalization error in this work 
(Thompson, priv. comm.), but agrees with a corrected expression given by
Thompson et al. (1999).

Using Eqn. (\ref{Edot_w}) with values for $P = P_{1806} = 7.48$ s,
$\dot P = \dot P_{1806} = 8.3\cdot 10^{-11}\ {\rm s}\ {\rm s}^{-1}$, 
i.e. those observed in SGR1806--20 and assuming the presence 
of a steady wind of luminosity $L_{37} = L_p/10^{37}$ erg s$^{-1}$, 
we obtain an estimate of the surface magnetic field of the neutron star:
\begin{equation}
B_0 \simeq  3 \times 10^{13}{\rm G} 
\left({P\over P_{1806}} \right)^{-1}    
\left({\dot P\over \dot P_{1806}} \right)
L_{37}^{-1/2},
\label{newB}
\end{equation}
where we have assumed $r_0 = 10$ km and $I = 10^{45}\,{\rm g}\,
{\rm cm}^2$. If one uses the observed value of the spin--down rate and 
the {\sl average} value of the particle luminosity needed to account 
for the energetics of the nebula, then $B_0$ is significantly below 
the estimate of $B_0 \simeq 10^{15}$
using Eqn (\ref{Edot_w}) with $R_{\rm LC} \simeq r_{\rm open}$.

This modified spin--down law leads to {\sl
exponential} increase in the pulsar period instead of the power law
increase associated with the purely dipole emission (this is easily
seen from equating $\dot E_w = -I\Omega\dot \Omega$ and integrating).
One can thus estimate
the age $\tau$ of SGR1806-20 through the relation
\be \label{tau_sw}
\tau = {P\over 2\dot P}\ln{\left[{L_p P^3\over 4\pi^2 I\dot P}\right]},
\ee
which yields $\tau\sim$ 11,800 yr for $L_p=10^{37}$~erg~s$^{-1}$.
Thus the steady wind model can naturally account for the fact that the age
of the SNR G10.0--0.3 is much larger than the characteristic dipole spin-down
age, but with the penalty that the magnetar model must be abandoned.

\section{COMBINED WIND AND DIPOLE SPIN-DOWN}

The expression for the magnetic field and characteristic age of the
neutron star given above are only valid if the relativistic particle 
wind completely and continuously dominates the spin down of the star.
If the particle wind flow is either discontinuous or not dominant,
then a more general description of the spin-down energy loss must be
used.  If the wind has instantaneous luminosity $L_p$ during its 
times of activity and duty cycle $D_p$, defined as the fraction of total
on-time, then the average energy loss from combined wind and
dipole is
\begin{eqnarray}
\dot E &=& -\langle I\Omega\dot \Omega \rangle 
= \dot E_D (1-D_p) + \dot E_w D_p \nonumber \\
&=& {B_0^2 r_0^6 \Omega^4 \over 6 c^3}\,(1-D_p) + L_p^{1/2} D_p { B_0 r_0^3 
\Omega^2 \over \sqrt{6 c^3}},
\label{Edot_gen}
\end{eqnarray}
where we have used Eqn (\ref{Edot_w}).
The surface magnetic field may then be found as the solution to the
quadratic equation, giving
\be \label{B0_gen}
B_0 = -{\sqrt{6c^3}\over 8\pi^2} {L_p^{1/2} D_p P^2\over (1-D_p) r_0^3} 
F(P,\dot P)
\ee
where,
\be
F(P,\dot P) = \left\{1 - \left(1+{4 \dot E (1-D_p)\over L_p 
D_p^2}\right)^{1/2}\right\},
\ee
and $\dot E = {4\pi^2 I\langle \dot P \rangle / P^3}$.
Note that when $L_p D_p^2 \ll 4 \dot E (1-D_p)$, 
\be
B_0 \simeq \left({3 c^3 I \langle \dot P\rangle P \over 2\pi^2 
r_0^6 (1-D_p)}\right)^{1/2}
\ee
which gives the standard dipole formula when $D_p = 0$.
If $L_p D_p^2 \gg 4 \dot E (1-D_p)$, 
Eqn (\ref{B0_gen}) gives the 
pure wind formula Eqn(\ref{newB}) with $L_p$ replaced by $L_p D_p^2$.

We may also integrate Eqn (\ref{Edot_gen}) from the initial period $P_0$ 
to the present period $P$ to obtain the general expression for the 
neutron star characteristic age $\tau$.  Assuming that $P_0 \ll P$,
\be \label{tau_gen}
\tau \simeq -{4 \pi^2 I \over L_p D_p^2 P^2}{\ln{[1-{2/F(P,\dot 
P)}]} \over F(P,\dot P)}.
\ee
This expression gives the usual characteristic age for dipole spin down,
$\tau = P/2\dot P$, in the limit $L_p D_p^2 \ll 4 \dot E (1-D_p)$ and
$D_p = 0$.  One must be careful in the limit $L_p D_p^2 \gg 4 \dot E (1-D_p)$.
If $D_p$ is close to 1, then the first term in Eqn (\ref{Edot_gen}) 
should be dropped to give an expression for $\tau$ which is the same as Eqn ({\ref{tau_sw}), again with $L_p$ replaced by $L_p D_p^2$.

We now examine the consequences of the general expressions 
(\ref{B0_gen}) and (\ref{tau_gen}) for SGR1806-20, the SGR source for which
we have the best estimate of the particle wind luminosity.  Figure 1 shows
the values of $B_0$ and $\tau$ computed from Eqns (\ref{B0_gen}) and 
(\ref{tau_gen}) for the measured $P $ and $\dot P $ of SGR 1806-20 
(Kouveliotou et al. 1998) as a function of the parameter $L_p D_p^2/ 
\dot E$, indicating the fractional wind contribution to the spin-down 
energy loss rate, assuming $D_p \ll 1$.  For small $L_p D_p^2/ \dot E$,
the curves approach their dipole radiation values of $B_0 \simeq 
10^{15}$ G and $\tau \simeq 1500$ yr.  For $L_p D_p^2/ \dot E \gsim 0.1$,
$B_0$ and $\tau$ begin to depart from these values, with the magnetic
field decreasing and the age increasing to connect smoothly to the 
wind-dominated solutions.  We have seen in Section II that assuming 
continuous, wind dominated spin-down can give a characteristic age 
that agrees with the age ($\tau \sim 10^4$ yr) of the plerion 
surrounding SGR1806-20, but that the derived magnetic field drops
into the range of normal radio pulsars.  In such a case, however, the 
free energy associated with the magnetic field decay is not sufficient to
account for the observed luminosity, and an alternative free energy
source must be considered.

The parameter $L_p D_p^2$ may be estimated for SGR 1806-20 from its observed
burst characteristics.  In general, we can write the particle luminosity
associated with a burst as $L_p = E_{\gamma}\,\epsilon_{\gamma}^{-1}\,
\Delta\tau_w^{-1}$,
where $E_{\gamma}$ is the $\gamma$-ray burst energy, $\epsilon_{\gamma}$ 
is the
conversion efficiency of particle energy to $\gamma$-rays and $\Delta\tau_w$
is the duration of the wind outflow following the burst.  If $T$ is the average 
time between bursts, then the wind duty cycle is $D_p = \Delta\tau_w/T$,
and
\be  \label{Lp} 
L_p D_p^2 = E_{\gamma}\,\epsilon_{\gamma}^{-1}\,\Delta\tau_w\,T^{-2}.  
\ee
In addition, the requirement that the X-ray nebula (Murakami et al. 1994) 
is powered by the aggregate of
the bursts' wind outflows leads to the condition $L_p D_p = 10^{42}\,{\rm erg}\,E_{40}(10^{-2}/\epsilon_{\gamma})/T = 10^{37}\,\rm erg\,s^{-1}\,(10^{-2}
/\eta)$, where $\eta$ is the conversion efficiency of particle 
luminosity to nebular emission and $E_{40} \equiv E_{\gamma}/10^{40}\,\rm erg$.
For the multiple SGR bursts from SGR 1806-20, Eqn (\ref{Lp}) and the above 
requirement on $L_p D_p$, we find $E_{40}\,(10^{-2}/\epsilon_{\gamma}) =
(T_{SGR}/10^5\,{\rm s})(10^{-2}/\eta)$, giving $L_p D_p^2 = 10^{34}\,{\rm erg\,s^{-1}}\,(\Delta\tau_w/10^2\,{\rm s})(10^5\,{\rm s}/T_{SGR})
(10^{-2}/\eta)$.  Likewise, for the giant bursts, we have $L_p D_p^2 = 3 \times
10^{35}\,{\rm erg\,s^{-1}}\,(\Delta\tau_w/10^7\,{\rm s})(10\,{\rm yr}/T_{G})
(10^{-2}/\eta)$.  
From Eqn (\ref{B0_gen}), Eqn (\ref{tau_gen}) and Figure 1, the conflicting goals of 
preserving the magnetar model (i.e. $B_0 \gsim 10^{14}$ G) and bringing the
characteristic age within a factor of 2 of the $10^4$ yr age of G10.0-0.3, may be
satisfied with $L_p D_p^2/ \dot E \simeq 10 - 100$.  Since $\dot E = 8 \times 10^{33}
\,\rm erg\,s^{-1}$, the duration of the particle outflow must be much larger than
the $\gamma$-ray burst duration and the wind flow duty cycle must be, 
$D_p \sim 0.008 - 0.08$. 

\section{DISCUSSION}

The results of our analysis leave two alternatives for interpreting the spin
down of SGRs, given the present limited data.  The first assumes a continuous
wind outflow at a luminosity sufficient to yield a characteristic
age in agreement with that of the surrounding SNR. We have shown
that in the case of SGR1806-20, this assumption leads to a surface 
magnetic field of only $B_0 = 3 \times 10^{13}$ G, well below the 
magnetar range ($>10^{14}$ G). However, this alternative requires 
a source of free energy other that the field decay to power both
the nebular emission and the SGR bursts.  The second option assumes an episodic
(or at least variable) wind outflow such that the average wind luminosity is
sufficient to power the nebular emission.  This allows a range of combinations of
surface field and age and depends on the wind duty cycle.  We show that, 
in the case of SGR1806-20, it is possible to accommodate both 
a characteristic age $\tau \sim 7500$ yr, consistent with the estimated 
SNR age of $\sim 10^4$ yr, and a magnetar model ($B_0 = 10^{14}$G).  
One should then observe
a sudden increase in the period derivative following SGR bursts.  Evidence for 
such an increase was seen following the bursting activity of June - August 1998 
from SGR1900+14 (Marsden et al. 1999).  These options assume
that gravitational radiation did not play a significant role in early spin-down 
evolution of the star, but it is unlikely to have made a large difference in the 
characteristic age.

There are a number of arguments in favor of the magnetar model for SGRs, most of 
which have been discussed by Thompson \& Duncan (1995) and Baring \& Harding (1998).
An additional argument is that the pure dipole fields of AXPs, which 
are spinning down smoothly (Gotthelf et al. 1999) and have much lower 
luminosity wind flows (if any), lie in the magnetar range.  
We suggest that detailed monitoring of the spin
periods of the SGRs, to search for variations in the period derivative,
can measure or place limits on the duty cycle of particle outflows and thus 
determine whether a magnetar model for these sources is viable.
\vskip 0.2 truecm
We thank Rob Duncan, Chris Thompson, Peter Woods and the referee Cole Miller
for valuable commments and discussions.

}

\newpage

\figureout{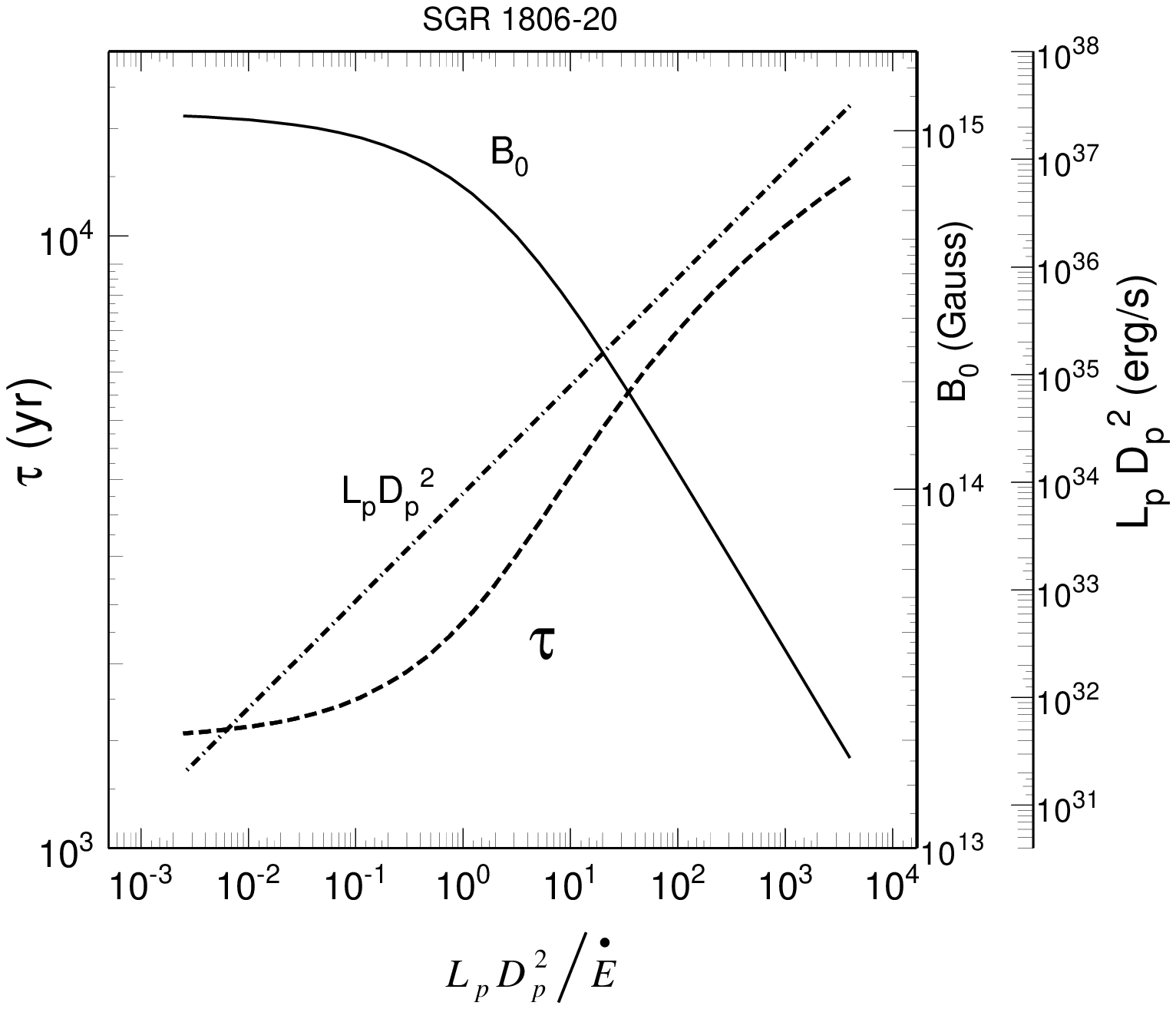}{7.5in}{0}
{Surface dipole magnetic field $B_0$ and characteristic age $\tau$
as a function of the parameter $L_p D_p^2/ \dot E$,
from the general expressions Eqns (\ref{B0_gen}) and (\ref{tau_gen}) derived from
the combined dipole and wind spin-down formula, Eqn (\ref{Edot_gen}).}

\end{document}